\documentclass[
aps,amsmath,superscriptaddress,nofootinbib]{revtex4}
\usepackage{graphicx}
\usepackage{slashed}

\usepackage{bm}
\usepackage{epsfig}

\newcommand{\bea}{\begin{eqnarray}}
\newcommand{\eea}{\end{eqnarray}}
\newcommand{\X}{X(3872)}
\newcommand{\Pb}{\bar{P}}
\newcommand{\Vb}{\bar{V}}
\newcommand{\nn}{\nonumber}

\date{30 April 2013}

\begin{document}


\title{Using the decay $\psi(4160) \rightarrow X(3872) \gamma$ to probe the molecular content of the $X(3872)$}

\author{Arman Margaryan}
\affiliation{Department of Physics, 
	Duke University, Durham,  
	NC 27708\vspace{0.2cm}}

\author{Roxanne P. Springer\footnote{Electronic address: rps@phy.duke.edu}}
\affiliation{Department of Physics, 
	Duke University, Durham,  
	NC 27708\vspace{0.2cm}}



\begin{abstract}
The $X(3872)$ has non-charmonium-like properties, such as decay processes
that seem to violate isospin, and a mass that lies unexpectedly close to the $D^0$ $\bar{D}^{0*}$ threshold. 
An EFT that includes both charmonium-like (short distance) and molecule-like
(meson bound state) properties is used to analyze the $X(3872)$ as it is
produced in the decay of $\psi(4160)$.  This is a route that BESIII may
be able to measure.  We find that the correlation between the angular distribution
of the outcoming photon (or $X(3872)$) and the polarization of the $\psi(4160)$ source
may be used to provide information on whether short-distance or long-distance
effects dominate.
\end{abstract}

\maketitle

\section{Introduction}

The $X(3872)$ was discovered by the Belle collaboration \cite{Choi:2003ue} as a narrow
 resonance from the decay $B^{\pm}\rightarrow X(3872)K^{\pm}$, $X(3872)\rightarrow J/\psi \pi^{+}\pi^{-}$. 
Its existence has been
confirmed by the CDF \cite{Acosta:2003zx},  D0 \cite{Abazov:2004kp}, and
BaBar \cite{Aubert:2004ns} collaborations, and now at the LHC \cite{Aaij:2011sn,Chatrchyan:2013cld}.  The most recent Particle Data Group value for
its mass is $m_{(X)}= 
3871.68 \pm 0.17$ \cite{pdg}, but whether it is actually above
or below the $D^0$ $\bar{D}^{0*}$ threshold at $3871.81\pm0.36$ MeV is still
an open question. The Belle collaboration finds an upper 
limit on the width of the $X(3872)$ to be $\Gamma_{(X)}<
1.2$ MeV at a 90 percent confident level \cite{Choi:2011fc}.

While uncertain for most of the time since its discovery, the $J^{PC}$ 
quantum number assignments for the X(3872) are now known to be
$1^{++}$ \cite{Aaij:2013zoa}.  This, along with the closeness of the $X(3872)$ to the $D^0$ $\bar{D}^{0*}$ threshold, makes it possible for the X(3872) to be interpreted
as a loosely bound state of $D^0$ and $\bar{D}^{0*}$ mesons.  The possibility that mesons could themselves form ``molecular" bound states of other mesons was discussed in Ref.~\cite{Weinberg:1962hj} and for charmed mesons in particular in Refs.\cite{Bander:1975fb,Voloshin:1976ap, De Rujula:1976qd,Nussinov:1976fg,Tornqvist:1993ng}.  The $\X$ was investigated as a potential molecule shortly after its discovery in Refs.\cite{Voloshin:2003nt,Close:2003sg,Swanson:2003tb,Tornqvist:2004qy,Braaten:2004fk}.   It seems certain that there is at least
a component of the $\X$ that can be taken as a molecule given that it will
likely strongly mix with the C=+1 combination of the neutral $D$ mesons.  Exactly
how much of it is molecular, what else might describe its wavefunction, and what observables should be studied to unravel it are the subject of lively debate in the literature.  

If the $X(3872)$ is indeed a molecule, then it is a very shallow bound state with a very large scattering length, possibly in excess of 6 fm.  This would make it larger than, for example, the deuteron.  The benefit of such a shallow bound state is that
its properties are dictated by this large scattering length.  The universal
properties of such systems is discussed in Ref.~\cite{Braaten:2004rn}.

In Ref.~\cite{Mehen:2011ds} we explored the behavior of the $X(3872)$ by noting that its production angular distribution depends upon the ratio of short-distance to long-distance terms.  In particular, using the X-EFT developed in Ref.~\cite{Fleming:2007rp} along with
heavy hadron chiral perturbation theory (HH$\chi$PT), we determined that the decay of the $\psi(4040) \rightarrow X(3872) \gamma$ depended upon diagrams that are dominated
by molecular-like (or long-distance) behavior and a single diagram that depends upon
a short-distance interaction.  Since this is an effective field theory (EFT) treatment, whether the short-distance operator mimics a $c \overline c$ character or some other short-distance character is not determined.  In particular, while the $\X$ may also mix with a linear combination of charged $D^{(*)}$ mesons as well as neutral ones, we consider those ``short-distance" (8 MeV above the $\X$ mass) on these scales.

In this paper we look at production of the $X(3872)$ from the decay 
$\psi(4160) \rightarrow X(3872) \gamma$.  BESIII intends to produce and 
study $\psi(4160)$ and in particular use it as a source of $X(3872)$ production \cite{Asner:2008nq}.  Like the $\psi(4040)$, the $\psi(4160)$ has quantum numbers 
$J^{PC}=1^{--}$ and is likely a traditional
charmonium excitation.  It is one of the $L=D$ multiplets, $2^3D_1$.  Its partial
fraction to electrons suggest that it may have additional $L$ admixtures, but since
this is uncertain at the moment we take it to be dominantly a pure state here. 
Its mass and width are estimated by the PDG to be $m_{\psi}=4153\pm3$ MeV and $\Gamma_{\psi}=103\pm8$ MeV respectively.


Below we find the differential cross section $\frac{d\sigma[\psi(4160)\rightarrow X(3872)\gamma]}{d\Omega}$ and extract its dependence on 
the angle between the outgoing photon momentum and initial $\psi(4160)$ polarization vector.  We discuss how this correlation can be used to determine the
short-distance versus long-distance character of the $X(3872)$.  
We also provide an estimate for the total decay rate $\Gamma[\psi(4160) \rightarrow \X \gamma]$ should the $\X$ be predominantly a bound state of neutral $D$ mesons,
${1 \over \sqrt{2}} \left( \bar D^0 D^{*0}+ \bar D^{*0}  D^0\right)$.

\medskip

\section{EFT Lagrangian}

To create an effective field theory for QCD, we identify the fields whose behavior
we want to describe, the energy region of interest, the symmetries we want to
impose,  and the small parameter that will organize the operators in the Lagrangian.
Then we write down 
the most general Lagrangian order by order \cite{Weinberg:1978kz}. 
In the limit $m_c \rightarrow \infty$ and $m_{d,u,s}\rightarrow 0$ QCD acquires two approximate symmetries: heavy quark spin 
symmetry and chiral symmetry \cite{16}. HH$\chi$PT is an effective field theory with both of those symmetries, including a simultaneous expansion in both limits.
The heavy hadrons are treated as nonrelativistic particles with their
classical mass term rotated away, leaving a derivative expansion in $p/m_c$, where  $p$ is the (small) momentum scale
in the problem \cite{16}. We will keep  the zeroth order terms in the chiral expansion, but include
the leading $p/m_c$ operator.  
 XEFT \cite{Fleming:2007rp} is an effective field theory describing low-energy nonrelativistic $D$, $\bar D$, $D^*$, $\bar{D}^*$,  and $\pi$ mesons near the $D^0+\bar{D}^{0*}$ mass threshold.  It is matched onto HH$\chi$PT by integrating out virtual states whose energies are  
widely separated from that  threshold.  It is similar to the NN-EFT created
to treat the deuteron as a bound state of nucleons \cite{Kaplan:1996xu}, but it is better behaved in
that pions can be treated perturbatively.  XEFT was designed to describe the $X(3872)$ as a bound state of $D^0/\bar D^{*0}$ + c.c. mesons.

For the $\psi(4160) \rightarrow \X \gamma$ decay we need HH$\chi$PT operators that include the 
 $\psi(4160)$, $D^{(*)0}$, and $\bar{D}^{(*)0}$ particles. 
The $D^{(*)}$ mesons are collected into a superfield to encode the heavy
quark symmetry. In the lowest multiplet the quarks $c \bar u$ form a bound state with relative orbital angular momentum $L=0$.  The quark spins combine to form
the  $J=0$ $D$ mesons (denoted $P$) and the $J=1$ $D^*$ mesons ($V^\mu$). In general this would include the charged and strange-ness containing
$D$ mesons as well, but here we only require the neutral ones.  The superfield is
\begin{equation}
H=\frac{1+\slashed{v}}{2} (V^\mu \gamma_\mu-P \gamma_5) \frac{1-\slashed{v}}{2} \ \ ,
\end{equation}
where $v^\mu$ is the heavy quark four-velocity.
Because the heavy hadrons are treated as static sources, there is no pair production.  The $\bar D^{(*)0}$ mesons have their own field, $\bar H$. 

General discussions about 
combining different spin and orbital angular momentum states into one field multiplet can be found in Refs.~\cite{17,18}.  Ref.~\cite{18} provides the multiplet fields for the
$\overline c c$ states within one $L$ value. 
Each of the quarks has spin $s=1/2$, so the $\bar c c$ state has spin $S=0$ or $S=1$. All the possible $J$ states with the same angular
 momentum $L$ are then given by $J=L$ when $S=0$, and $J=|L-1|,L,L+1$ when 
$S=1$. The particle $\psi(4160)$ consists of $c \bar c$ quarks which have relative $L=2$ angular momentum. So the field multiplet in which $\psi$(4160) lives is \cite{18}:

\begin{eqnarray} \label{jmunu}
J^{\mu \nu} &=& \frac{1+\slashed{v}}{2}(H_3^{\mu \nu \alpha} \gamma_\alpha + \frac{1}{\sqrt6}(\epsilon^{\mu \alpha \beta \gamma} v_\alpha \gamma_\beta H^\nu_{2 \gamma}+\epsilon^{\nu \alpha \beta \gamma} v_\alpha \gamma_\beta H^\mu_{2 \gamma})     \nonumber \\
   &+& \frac{1}{2}\sqrt{\frac{3}{5}}((\gamma^\mu-v^\mu)H^\nu_1+(\gamma^\nu-v^\nu)H^\mu_1) \nonumber \\
   &-& \frac{1}{\sqrt{15}}(g^{\mu \nu}-v^\mu v^\nu) \gamma_\alpha H_1^\alpha+K_2^{\mu \nu} \gamma_5) \frac{1-\slashed{v}}{2} \ \ ,
\end{eqnarray}
where  $H_A$, $K_A$ are the effective fields of the various members of the multiplet with total spin $J=A$.
 Since the total spin of $\psi(4160)$ is $J=1$
 we need only the A=1 term.

Using HH$\chi$PT power counting we identify the leading order operators that couple $D^{(*)}$ mesons
to photons, the $\psi(4160)$ to the $D^{(*)}$ mesons, and the $\psi(4160)$ to both 
$D^{(*)}$ mesons and photons.

\begin{eqnarray} \label{lag}
\mathcal{L} &=& \frac{e \beta}{2}  Tr( H^\dag H \vec{\sigma} \cdot \vec{B} Q)+\frac{eQ^{'}}{2m_c}Tr(H^\dag  \vec{\sigma} \cdot \vec{B} H) \nn \\
&+&  i\frac{g}{2}Tr (J^{ij} \bar {H}^\dag \sigma_i \stackrel{\leftrightarrow}{\partial_j} H^\dag)
+i\frac{ec}{2}Tr(J^{ij} J \sigma_i E_j \bar H)+ h.c. \ \ , 
\end{eqnarray}
where we use the 2-component notation of Ref.~\cite{Hu:2005gf}, with
\bea
H=\vec V \cdot \sigma + P 
\eea
the superfield that contains both the vector $V_i$ field of the $D^{*0}$
and the pseudoscalar field $P$ of the $D^0$.  Because we are confining
ourselves to the neutral D-mesons only, $Q$=2/3, and the isospin subscripts are dropped.  $\sigma_j$ is the spin Pauli matrix.  The second term in Eq.~(\ref{lag}) contains the coupling to the charm quark of $Q^\prime=2/3$.  For a discussion of these and higher order EM couplings among the $D$-mesons and their excited
states see Ref.~\cite{Stewart:1998ke}.  In the nonrelativistic limit our $\psi(4160)$ 
(the $H_1$ of Eq.~(\ref{jmunu}))is now called $\psi^i$,
\begin{equation}
J^{i j}= \frac{1}{2}\sqrt{\frac{3}{5}}(\sigma^i \psi^j+\sigma^j \psi^i)-\frac{1}{\sqrt{15}} \delta^{i j} \sigma^\alpha \psi^\alpha
\end{equation}
The coefficients $\beta$, $g$, and $c$ will be discussed in the ``Estimates of Parameters" section.

\section{The decay $\psi(4160) \rightarrow \X \gamma$}

The Feynman diagrams contributing to the decay are given in Fig.~1.
\begin{figure}[ht]
  \begin{center}
  \includegraphics[scale=.75]{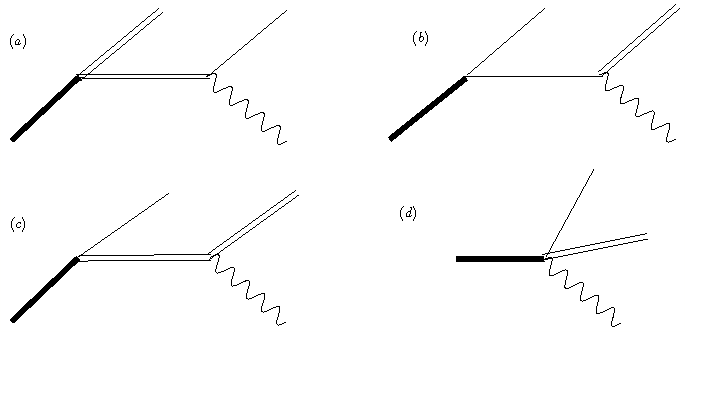}
  \end{center}
  \caption{Feynman diagrams contributing to the decay $\psi(4160) \rightarrow D^0 \bar D^{*0}$. The thick solid line is the $\psi(4160)$ particle, the thin solid line is a $D^0$ particle,
the double line is a $\bar{D}^{0*}$ particle and the wavy line is a photon.}
\end{figure}
Using the rules obtained from Eq.~(\ref{lag}) we find each contributes the
following amplitude: \footnote{These were also calculated by T. Mehen (unpublished).}

\begin{eqnarray}
(a)&=&-\frac{g \beta_+ e}{3 \sqrt{15}} \frac{1}{E_\gamma-\Delta} (4(\vec{\epsilon}_\psi \cdot \vec{k}) (\vec{\epsilon}_{D^*} \cdot  \vec{k} \times \vec{\epsilon}^{\: *}_\gamma)-
(\vec{k} \cdot \vec{\epsilon}_{D^{*}})(\vec{\epsilon}_\psi \cdot \vec{k} \times \vec{\epsilon}^{\: *}_\gamma) \\
(b)&=&-\sqrt { \frac{5}{3} } \frac{2}{3} \frac{g e \beta_{+}} {E_{\gamma} + \Delta} ( \vec{k} \cdot \vec{\epsilon}_{\psi}) 
( \vec{\epsilon}_{D^{*}} \cdot \vec{k} \times \vec{\epsilon}^{\: *}_{\gamma}) \\
(c)&=& \frac{1}{3} \sqrt{\frac{5}{3}}\beta_-\frac{ge}{E_\gamma}(\vec{k} \cdot \vec{\epsilon}_{D^{*}})(\vec{\epsilon}_\psi \cdot \vec{k} \times \vec{\epsilon}^{\: *}_\gamma) \\
(d)&=&-\frac{1}{2}\sqrt{\frac{5}{3}} e c E_\gamma \vec{\epsilon}_{D^{*}} \cdot \vec{\epsilon}_\psi \times \vec{\epsilon }^{\: *}_\gamma \ \ ,
\end{eqnarray}
where $\beta_\pm=\beta \pm \frac{1}{m_c}$; the polarization vectors of the photon, $D^{0*}$, and $\psi(4160)$ are $\vec{\epsilon}_{\gamma}$, $\vec{\epsilon}_{D^*}$ 
and $\vec{\epsilon}_\psi$, respectively; and $\vec{k}$ is the outgoing photon momentum. 
The decay rate depending on the polarization of the  initial $\psi(4160)$ is found by summing over the final photon and $D^{0*}$ particle polarizations:
\begin{equation}
\Gamma (\vec\epsilon_\psi)\sim \frac{2}{3}(A+C)^2 |\hat{k}\cdot\vec{\epsilon}_\psi|^2+\frac{1}{3}(B-C)^2 |\hat{k} \times \vec{\epsilon}_\psi|^2
\end{equation}
where $\hat{k}$ is the unit vector in the direction of the photon's 3-momentum, and
\begin{equation} \label{A}
A=g\beta_+ e \frac{2}{3\sqrt{15}}E_\gamma^2 \frac{7E\gamma-3\Delta}{\Delta^2-E_\gamma^2} \ \ ,
\end{equation}
\begin{equation} \label{B}
B=\frac{ge}{3\sqrt{15}} \frac{-\beta_+ E_\gamma^2+5\beta_-E_\gamma(\Delta-E_\gamma)}{\Delta-E_\gamma} \ \ ,
\end{equation}
\begin{equation} \label{C}
C=-ecE_\gamma \frac{1}{2} \sqrt{\frac{5}{3}} \ \ .
\end{equation}
Averaging this over $\psi(4160)$ polarizations gives the total decay rate
\begin{equation}
\Gamma \sim \frac{2}{3}((A+C)^2+(B-C)^2) \ \ .
\end{equation}
If the $\psi(4160)$ is produced in an electron-positron collider such as BESIII
then it is produced with a polarization normal to the beam axis in the limit
that the electrons can be treated as massless helicity eigenstates.  So we can
use the relationship between the outgoing photon  (or ultimately $\X$) momentum $\vec k$ in the $\psi(4160)$ rest frame with
respect to $\vec \epsilon_\psi$ to obtain a relationship between $\vec k$ and the beam axis.  

Defining
\begin{equation}
P=\frac{2}{3}(A+C)^2                      
\end{equation}
and
\begin{equation}
T=\frac{2}{3}(B-C)^2 \ \ ,
\end{equation}
the angular distribution of the final states is
\begin{equation}
\frac{d\Gamma}{d {\rm cos}\theta} \sim 1+\rho \ {\rm cos}^2\theta   \ \ ,\hspace{1.5cm} \rho=\frac{T-2P}{T+2P} \ \ ,
\end{equation}
where $\theta$ is the angle between the photon momentum vector and $\psi(4160)$ polarization.
Substituting expressions for A, B and C yields
\begin{equation}
\rho=\frac{(\frac{2}{15}\frac{-E_\gamma+5r_\beta(\Delta-E_\gamma)}{\Delta-E_\gamma}+\eta)^2-2(\frac{4}{15}E_\gamma\frac{7E_\gamma-3\Delta}{\Delta^2-E_\gamma^2}-\eta)^2}{(\frac{2}{15}\frac{-E_\gamma+5r_\beta(\Delta-E_\gamma)}{\Delta-E_\gamma}+\eta)^2+2(\frac{4}{15}E_\gamma\frac{7E_\gamma-3\Delta}{\Delta^2-E_\gamma^2}-\eta)^2} \ \ ,
\end{equation}
where $r_\beta=\beta_-/\beta_+$ and $\eta=\frac{c}{g\beta_+}$.
  This ratio $\eta$ provides a measure of how much of the decay behavior and
  polarization correlation is driven by the long-distance diagrams (Fig. 1(a)-1(c))
  that depend upon $g \beta_\pm$ versus the short-
 distance contact $c$-dependent diagram in Fig. 1(d). The plot of $\rho$ as a function of $\lambda$ is given in Fig. 2.  Varying $r_\beta$ within reasonable ranges does not change the shape of this curve.  If the $\psi(4160) \rightarrow X(3872) \gamma$ decay were driven entirely by long-distance physics then $\eta=0$ and  $\rho\sim-0.8$ (for $r_\beta=1$) or $\rho\sim-0.9$ (for $r_\beta=0.66$).  A measurement of 
  $\rho \sim -1/3$ would not be definitive because that is supported by either 
  $\eta \rightarrow \infty$ (short-distance dominance) or  $\eta \sim -1.5$. But
  finding $\rho \le -0.7$ or $\rho > 0$ would suggest a significant long-distance contribution to the decay. 
  \begin{figure}[h]
  \begin{center}
  \includegraphics[scale=.75]{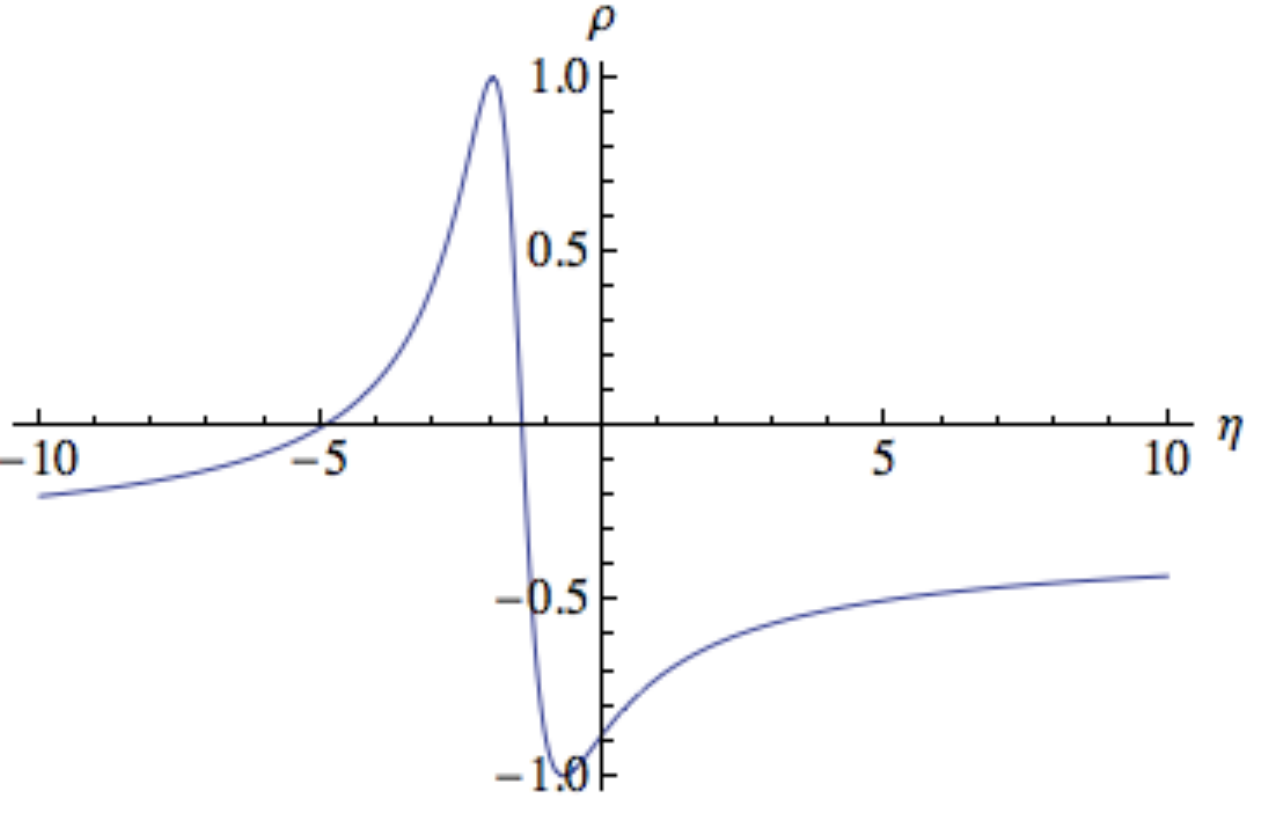}
  \end{center}
  \caption{ $\rho$ as a function of parameter $\eta$, with $r_{\beta}=0.66$.}
\end{figure}

XEFT is used to match the $\psi(4160)$ decays in Fig. 1 to $\psi(4160) \rightarrow
\X \gamma$, but this just provides an overall constant that cancels in the $\rho$ parameter

\section{Estimates of parameters}

In this section we discuss the parameters $\beta$, $g$, and $c$ from Eq.~(\ref{lag}) as well as the unknown matrix element between the $\X$ and the constituents
$1/\sqrt{2} \left( \bar D^0 D^{*0}+ \bar D^{*0}  D^0\right)$.  

The coefficient $\beta$ in Eq.~(\ref{lag}) is found from measured electromagnetic decays
among the $D$ mesons and their excited states.  Ref.~\cite{Hu:2005gf} finds $\beta^{-1} = 275-375$ MeV 
using conditions similar to the ones relevant for this calculation.

Now we will use some experimental limits and theoretical estimates to provide
an order of magnitude expectation for the branching fraction expected for $\psi(4160) \rightarrow \X \gamma.$  
The partial width found in Ref.~\cite{Aubert:2008ae} (but note that Belle does not see this decay \cite{Bhardwaj:2011dj}) is
\begin{eqnarray}
{ \Gamma[X(3872) \rightarrow \psi(2S) \gamma] \over \Gamma_{tot}}> 0.03 \ \ .
 \end{eqnarray}
 A lower limit on the total width of the $X(3872)$ is estimated to be the width of
 the $D^*$, or about 70 keV.  This provides a lower limit 
 $\Gamma[X(3872) \rightarrow \psi(2S) \gamma]  > 0.002 $ MeV.
   In Ref.~\cite{Mehen:2011ds} we found that
 \bea\label{tr}
\Gamma[\X \rightarrow \psi(2S) \gamma] &=& 
\sum_\lambda |\langle 0| \frac{1}{\sqrt{2}}{\epsilon}_i(\lambda) 
\,(V^i \, \Pb +\Vb^i \, P) |X(3872,\lambda)\rangle|^2\nn \\
&&\times \frac{E^{(2S)}_\gamma}{36\pi}  \frac{m_{\psi(2S)}}{m_X} \left[(A_{2}+C_{2})^2+(B_{2}-C_{2})^2\right]  \ \ ,
\eea
where we have replaced the $A$, $B$, and $C$ coefficients in Ref.~\cite{Mehen:2011ds} with $A_{2}$, etc., so they won't  be confused with the $A$, etc. in Eqs.~(\ref{A})--(\ref{C}) above.

The first term in Eq.~(\ref{tr}) is the matrix element $|{\cal M}|^2$ that 
encodes the overlap between the $X(3872)$ and the constituents $V^i$ (the vector mesons $D^*$), $P$ (the meson D), etc.  This matrix element is not known (although it could be estimated using effective range theory were the binding energy well measured) but it appears in all $\X$ production/decay cross sections.  If we are able to extract it from one measurement, we can then use it in predictions for others. 
For example,
\bea
\Gamma[\X \rightarrow \psi(2S) \gamma] &=& |{\cal M}|^2 F_1(m_{\psi(2S)},m_X,A_2,B_2,C_2) \nonumber \\ 
\Gamma[\psi(4160) \rightarrow X(3872) \gamma] & = & |{\cal M}|^2 F_2(m_{4160},m_X,A,B,C) \ ,
\eea
where $|{\cal M}|^2$ is universal and $F_1$  and $F_2$ are known functions of the parameters.

 The
coefficients $A_{2}$ and $B_{2}$ depend upon the coupling between the $\psi(2S)$ and 
the $D$, $D^*$, etc. mesons.  The coefficient $C_{2}$ depends upon an unknown short-distance constant.  Refs.~\cite{Guo:2009wr, Guo:2010zk} have estimated the coupling between $\psi(2S)$ and $D$ mesons to be $g_2 \sim 2$ GeV$^{-3/2}$.  (This is the same as the $g_2^\prime$ of Ref.~\cite{Mehen:2011tp}, which finds a value as low as 0.55 GeV$^{-3/2}$.)  This $g_2$ is the analog of the $g$
coupling in the Lagrangian of Eq.~(\ref{lag}) that couples $\psi(4160)$ to $D$ mesons. If we assume that indeed the $X(3872)$ is dominated by a molecular configuration such 
that we can neglect the impact of $C_2$, we find that  experimental
limits provide a limit on the matrix element squared above to be $|{\cal M}|^2>0.005$ GeV$^3$.

To estimate the rate of $\psi(4160) \rightarrow X(3872) \gamma$ 
still requires that we estimate the $g$ in Eq.~(\ref{lag}).  This can be attempted by comparing the partial widths of the $\psi(4160)$ to $D^{(*)}+D^{(*)}$ estimates in the quark model from Ref.~\cite{Eichten:2005ga}.  We find that $g\sim 1$ GeV$^{-3/2}$.  Collecting these, and again
assuming that the $\X$ behavior is dominated by long-distance physics, we
can give an order of magnitude estimate that the decay rate 
\bea
\Gamma[\psi(4160) \rightarrow \X \gamma] > 1 \ {\rm keV} \ \ ,
\eea
or a branching fraction of greater than 10$^{-5}$. 
The parameter $c$ might be estimated by saturating with nearby intermediate states, but we see that we can learn something about the $\X$ even without specific knowledge of $c$.  


\section{Summary}
We have discussed the differential cross section for $\psi(4160) \rightarrow \X \gamma$ by assuming that it has a nonzero overlap with a ``molecular" bound
state ${1\over \sqrt{2}} \left( \bar D^0 D^{*0} + \bar D^{*0}  D^0\right)$.  We argue
that a measurement of the angular distribution of the photon (or $\X$) with
respect to the beam axis can provide information on whether short-distance
(charmonium-like) or long-distance (molecule-like) behavior dominates in this decay.  
We have also provided an estimate for the branching fraction of the decay
route.

\pagebreak

\centerline{ACKNOWLEDGMENTS}
We thank T. Mehen for discussions, an independent calculation of the amplitudes, and for reading the manuscript.  We thank C. Hanhart and F.-K. Guo for bringing an error to our attention. 
This work was supported in part by the Department of Energy grant DE-FG02-05ER41368.
RPS thanks the Department of Energy's Institute for Nuclear Theory at the University of Washington for its hospitality while some of this work was completed.

\end{document}